\documentclass[twocolumn,showpacs,showkeys,amsmath,amssymb,superscriptaddress,floatfix,nofootinbib]{revtex4-1}

\usepackage{graphicx}
\usepackage{bm} 
\usepackage{subfigure}

\newcommand{\rb}[1]{\mbox{\textrm{\scriptsize #1}}}

\newcommand*{\roots}{\ensuremath{\sqrt{s_{_{NN}}}}}

\newcommand{\CuAucol}{\mbox{${}^{63}{\rm Cu}$+${}^{197}{\rm Au}$}}
\newcommand{\UUcol}{\mbox{${}^{238}{\rm U}+{}^{238}{\rm U}$}}
\newcommand{\nucCu}{\mbox{${}^{63}{\rm Cu}$}}
\newcommand{\nucAu}{\mbox{${}^{197}{\rm Au}$}}
\newcommand{\nucU}{\mbox{${}^{238}{\rm U}$}}

\makeatletter
\newcommand\erfc{\mathop{\operator@font erfc}\nolimits}
\def\slashchar#1{\setbox0=\hbox{$#1$}
   \dimen0=\wd0 \setbox1=\hbox{/} \dimen1=\wd1
   \ifdim\dimen0>\dimen1 \rlap{\hbox to \dimen0{\hfil/\hfil}} #1
   \else  \rlap{\hbox to \dimen1{\hfil$#1$\hfil}} / \fi}

\begin{document}
 
\title{Influence of initial fluctuations on geometry measures in relativistic U-U and Cu-Au collisions}

\author{Maciej Rybczy\'nski} 
\email{Maciej.Rybczynski@ujk.edu.pl}
\affiliation{Institute of Physics, Jan Kochanowski University, PL-25406~Kielce, Poland} 

\author{Wojciech Broniowski} 
\email{Wojciech.Broniowski@ifj.edu.pl}
\affiliation{Institute of Physics, Jan Kochanowski University, PL-25406~Kielce, Poland} 
\affiliation{The H. Niewodnicza\'nski Institute of Nuclear Physics, Polish Academy of Sciences, PL-31342 Krak\'ow, Poland}

\author{Grzegorz Stefanek} 
\email{Grzegorz.Stefanek@ujk.edu.pl}
\affiliation{Institute of Physics, Jan Kochanowski University, PL-25406~Kielce, Poland}

\date{12 November 2012}

\begin{abstract}
In the framework of the Glauber approach we investigate the influence of the initial fluctuations on various measures of the initial-state 
geometry in \CuAucol~and \UUcol~relativistic ion collisions. Comparing variants of Glauber model (the wounded-nucleon model, the mixed model, and the 
hot-spot model) we indicate sensitivity of certain observables, in particular for the azimuthal eccentricity parameters as well as 
for the correlation of directions of the principal axes associated with the Fourier components. We apply {\tt GLISSANDO} in our analysis.
\end{abstract}

\pacs{25.6.-q, 25.6.Dw, 25.6.Ld}


\maketitle 

\section{Introduction}
\label{sect:intro}

Variants of the Glauber model~\cite{Glauber:1959aa,Czyz:1969jg}, in particular the wounded-nucleon model~\cite{Bialas:1976ed,Bialas:2008zza} and its extensions~\cite{Kharzeev:2000ph,Broniowski:2007nz} have become a basic tool in modeling the early stage of relativistic heavy-ion collisions. An alternative treatment is based on the Color Glass Condensate theory (for a recent overview see, e.g.,~\cite{Gelis:2010nm} and references therein). The Glauber model approach provides initial conditions for the subsequent hydrodynamic evolution. That way the features of the nuclei structure, such as distributions of nucleons in nuclei, the NN correlations~\cite{Broniowski:2010jd}, as well as the NN cross section~\cite{Rybczynski:2011wv} show up indirectly in the measured observables.
Moreover, as argued by Filip et al.~\cite{Filip:2007tj,Filip:2009zz,Filip:2010zz}, the nucleus deformation plays an important role in the ``geometry'' of the collision. Since recently collisions \UUcol~ were registered at BNL RHIC, the issue of a proper inclusion of the nuclear deformation is important. Similarly, the measured collisions of asymmetric nuclei, \CuAucol, provide valuable information of the initial state geometry of such systems~\cite{Bozek:2012hy}, which lead the geometric triangular flow~\cite{Alver:2010gr,Alver:2010dn,Petersen:2010cw}. 

Many elements enter the modeling of the relativistic heavy-ion collisions: nuclear structure, models of the early phase, hydrodynamics, hadronization, rescattering of hadrons after freeze-out. Each of them brings in certain assumptions affecting the predictions for the observed quantities. As one is primarily interested in properties of matter created in the collision and inferred from the hydrodynamic phase, the early ``geometric'' phase should be modeled as accurately as possible to limit the uncertainty in the later stages. 
In particular, the effects of the deformation should be incorporated for such systems as the \UUcol~ 
collisions, since they may lead to observable effects in the elliptic flow in most central collisions~\cite{Voloshin:2010ut}. 
Asymmetric collisions, such as \CuAucol, differ qualitatively from the symmetric case, since they give rise
to the odd collective flow components not only from fluctuations, but also from the averaged original 
geometry of the collision, which in this case contains the odd Fourier components in the azimuth. 

Many recent analyses are devoted to the studies of the initial geometry and its influence for the further stages of the evolution and the observed quantities, both in approaches involving the Glauber model  \cite{Broniowski:2007ft,Andrade:2009em,Staig:2010pn,Teaney:2010vd,Qin:2010pf,Nagle:2010zk,Xu:2010du,Lacey:2010av,Werner:2010aa,Qin:2011uw,Bhalerao:2011yg,%
Bhalerao:2011bp,Gardim:2011xv,Qiu:2011hf,Qiu:2011iv,Xu:2011jm,Teaney:2012ke,Jia:2012ma,Hirano:2012kj} and the glasma \cite{Qiu:2011iv,Albacete:2011fw,Muller:2011bb,Gavin:2011gr,Heinz:2011mh,Schenke:2012wb,Schenke:2012fw,Dumitru:2012yr,Gavin:2012if}.
Studies of correlations between the harmonic flow components were presented in~\cite{Nagle:2010zk,Lacey:2010av,Bhalerao:2011yg,Jia:2012ma}.
We note that the experiment shows a strong correlation between $\Phi_2$ and $\Phi_4$ \cite{Adams:2003zg,Adare:2010ux}, while a weak correlation between $\Phi_2$ and $\Phi_3$ is found \cite{ALICE:2011ab}, where $\Phi_n$ denotes the principal axis associated with the $n$-th harmonic flow~\cite{Broniowski:2007ft}. The early phase of the \UUcol~ collisions has been investigated in the Glauber approach~\cite{Heinz:2004ir,Kuhlman:2005ts,Masui:2009qk,Hirano:2010jg}
as well as in the in AMPT model~\cite{Haque:2011aa}.

Notably, the generalization of the approach in the form of the mixed model~\cite{Kharzeev:2000ph,Back:2001xy,Back:2004dy}, amending the wounded nucleon model with some binary collisions, leads to successful description of centrality dependence of multiplicities in collisions at energy range from RHIC to LHC. This means that we have comparable average entropy production from ``soft'' wounded nucleons and from ``semi-hard'' binary collisions. In this model the production is proportional to $(1-\alpha)N_{\rb{W}}/2+\alpha N_{\rb{bin}}$, with $N_{\rb{W}}$ denoting the number of wounded nucleons, $N_{\rb{bin}}$ the 
number of binary collisions, and the parameter $\alpha$ controlling their relative weight.

However, it should be stressed that there is no ``unique Glauber model'' of the early phase. The original distribution of sources (wounded nucleons, binary collisions) should be overlaid with some properly chosen distribution of the entropy or energy production occurring at each individual NN collision. The idea goes back to the very beginnings of the wounded nucleon model~\cite{Bialas:1976ed}, where a wounded nucleon produces particles with a given distribution of the average multiplicity such that the measured hadron multiplicity is reproduced.
Adjusting the dispersion of this overlaid distribution one might also reproduce the multiplicity fluctuations in the nucleus-nucleus collisions. Depending on the form of the overlaid distribution, large fluctuations of the spatial entropy distribution may be induced. They lead to large fluctuations of the initial geometry, much larger compared to the naive case where no distribution is overlaid. These effects are studied in detail in this work in the context of the considered collisions.

A physically motivated overlaid distribution with large fluctuations is present in the {\em hot-spot} model, constructed in the spirit of~\cite{Gyulassy:1996br}. The model assumes that the cross-section for a semi-hard binary collision producing a hot-spot is small, around $\sigma_{\rm bin} \simeq 2$~mb, much smaller than the cross-section for the wounding, $\sigma_{\rm w} \simeq 40-60$~mb. However, when this rare collision occurs, it produces a very large amount of entropy, enhanced by the factor \mbox{$\sigma_{\rm w}/\sigma_{\rm bin}$} compared to the production from a wounded nucleon. This factor is chosen such way that the fraction of the production from the binary collisions is equal $(1-\alpha)$ by construction.

On top of the wounded nucleons and the hot-spots one may still overlay a suitable statistical distribution of the strength of the entropy production. Here, following~\cite{Broniowski:2007ft}, we use the $\Gamma$ distribution, which additionally increases fluctuations. The presence of hot-spots and the overlaid distribution, while from construction innocuous for the quantities dependent on the averaged densities, affects significantly the fluctuations, and consequently, the fluctuation measures, but also the harmonic flow measures. The paper is precisely focused on this class of effects. We study, with the help of {\tt GLISSANDO}~\cite{Broniowski:2007nz}, the following aspects of the recently measured Cu+Au and U+U reactions at RHIC:

\begin{table}

\caption{The parameters of the Woods-Saxon potential and deformation coefficients for the nuclei used in our analysis, taken from~\cite{Moller:1993ed}.\label{tab:nuclpar}}

{\begin{tabular}{ccccc} \toprule
                  &          &           &             & \\ 
nucleus           & $R$~[fm] & $a$~[fm]  & $\beta_{2}$ & $\beta_{4}$ \\ 
                  &          &           &             &  \\  \colrule
 \nucCu           & 4.206    & 0.5977    & 0.162       & -0.006 \\ 
 \nucAu           & 6.430    & 0.45      & -0.13       & -0.03  \\ 
 \nucU            & 6.810    & 0.54      & 0.28        & 0.093  \\ 
            &          &           &             & \\  \botrule
\end{tabular}
}

\end{table}

\begin{table}
\caption{Centrality classes (ranges in RDS) for the Cu+Au collisions
\label{tab:cent_CuAu}}

{\begin{tabular}{cccc} \toprule
                  &               &                &               \\ 
Centrality [\%]   & Wounded       & Mixed          & Hot-spot      \\ 
                  &               &                &               \\  \colrule
 0 - 5            & $>$ 101         &  $>$ 157.4       & $>$ 157.7 \\ 
 5 - 10           & 101 - 87  & 157.4 - 128.3  & 157.7 - 128.2 \\ 
 10 - 20          & 86 - 61   & 128.3 - 84.0   & 128.2 - 84.0  \\ 
 20 - 30          & 60 - 42   & 84.0 - 53.7    & 84.0 - 53.5   \\
 30 - 40          & 41 - 28   & 53.7 - 33.2    & 53.5 - 32.9   \\
 40 - 50          & 27 - 17   & 33.2 - 19.6    & 32.9 - 19.7   \\
 50 - 60          & 16 - 10    & 19.6 - 11.3    & 19.7 - 11.3   \\
 60 - 70          & 9 - 6     & 11.3 - 6.2     & 11.3 - 5.9    \\
 70 - 80          & 5 - 4     & 6.2 - 3.4      & 5.9 - 3.0     \\
                  &               &                &               \\  \botrule
\end{tabular}
}

\end{table} 

\begin{table}
\caption{Centrality classes (ranges in RDS) for the U+U collisions
\label{tab:cent_UU}}

{\begin{tabular}{cccc} \toprule
                  &                &                &                \\ 
Centrality [\%]   & Wounded        & Mixed          & Hot-spot       \\ 
                  &                &                &                \\  \colrule
 0 - 5            & $>$ 196  & $>$ 331.3  & $>$ 331.9  \\ 
 5 - 10           & 196 - 168  & 331.3 - 272.4  & 331.9 - 272.3  \\ 
 10 - 20          & 167 - 120  & 272.4 - 182.9  & 272.3 - 182.8  \\ 
 20 - 30          & 119 - 84   & 182.9 - 119.3  & 182.8 - 119.3  \\
 30 - 40          & 83 - 56    & 119.3 - 75.4   & 119.3 - 75.0   \\
 40 - 50          & 55 - 36    & 75.4 - 44.8    & 75.0 - 44.2    \\
 50 - 60          & 35 - 22    & 44.8 - 24.4    & 44.2 - 25.0    \\
 60 - 70          & 21 - 11    & 24.4 - 12.6    & 25.0 - 12.5    \\
 70 - 80          & 10 - 6     & 12.6 - 5.5     & 12.5 - 5.7     \\
                  &                &                &                \\  \botrule
\end{tabular}
}

\end{table}

\begin{itemize}

\item Sensitivity of the flow coefficients $\epsilon_{n}$ to the increased initial fluctuations of the entropy production (the hot-spot + $\Gamma$ model), as well as dependence on the nuclear deformation. Interestingly, we find that the fluctuations from hot-spots completely wash-out the knee structure for the ultra-central U+U collisions 
advocated in~\cite{Voloshin:2010ut}. 

\item Correlations of the harmonic principal axes and their sensitivity to the details of the initial model for the Cu+Au and U+U reactions, with the conclusion that the hot-spots change the qualitative behavior of these correlations. The effect is seen in the distributions of the differences of principal axes.

\item In our simulations we use the realistic NN collision profile. This profile is related to the differential NN cross section~\cite{Bialas:2006qf}. We have shown previously that the single Gaussian wounding profile is sufficiently accurate and is the physical choice~\cite{Rybczynski:2011wv}, differing 
in predictions for the fluctuation measures from the commonly used but inappropriate hard-sphere wounding profile.
 
\end{itemize}

\section{Model}
\label{sect:model}

\begin{figure*}
\begin{center} 
\includegraphics[width=1.3\textwidth]{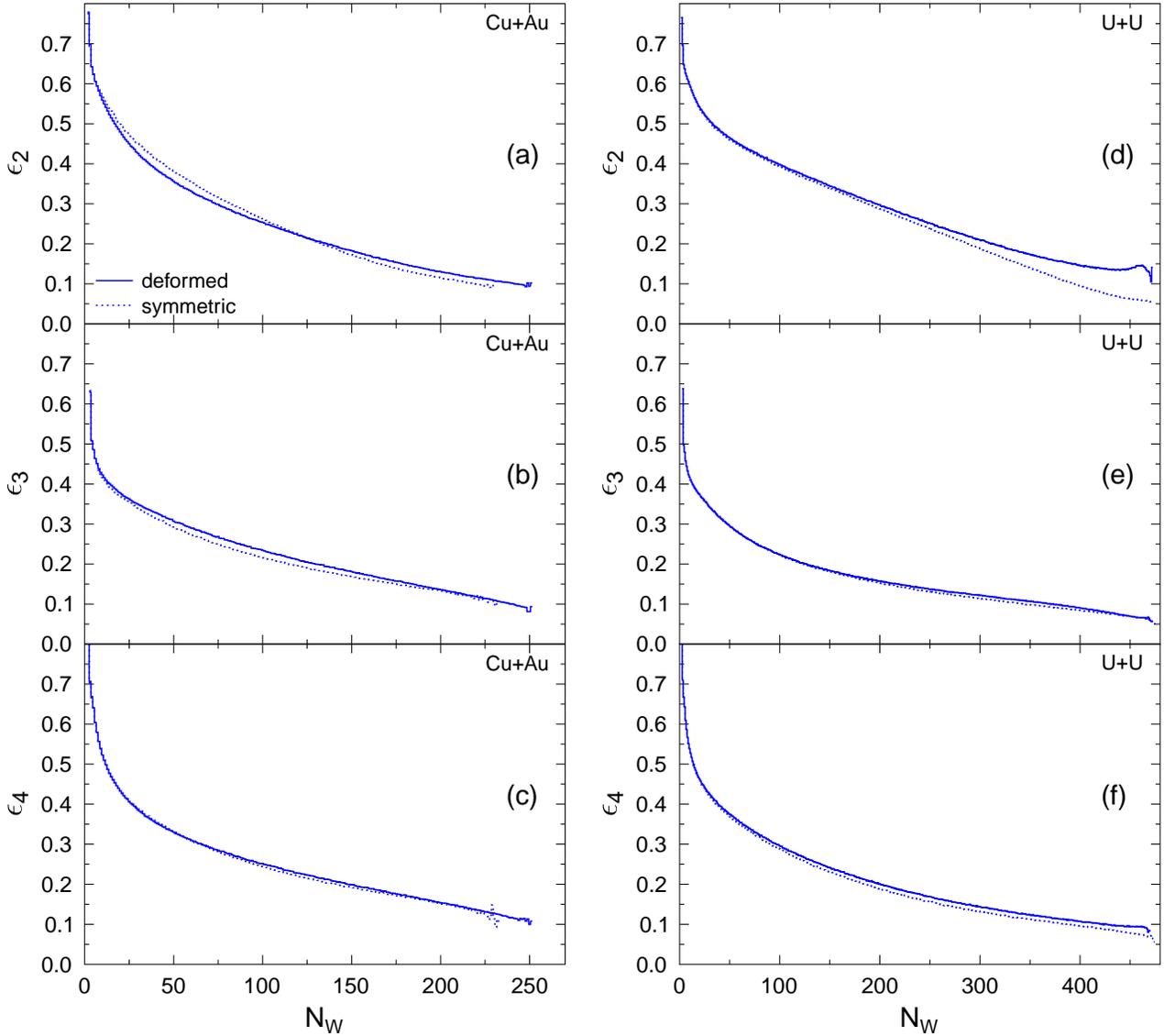}
\end{center}
\caption{(Color online) The event-average harmonic eccentricities $\epsilon_n$ ($n=2,3,4$) for the Cu+Au and U+U collisions, plotted as function of the number of wounded nucleons, $N_{\rm W}$. Mixed model, $\sigma_{\rm w}=42$~mb with the Gaussian wounding profile. The effects of the nuclear deformation are relevant for $\epsilon_2$ for the most central U+U collisions. 
\label{fig_epsn_mixed_gwp_rn_comp}} 
\end{figure*}

In the analysis presented in this paper we use {\tt GLISSANDO}~\cite{Broniowski:2007nz} modified to incorporate the 
shape deformation of nuclei. The spatial distribution of nucleons has been generated according to the deformed Woods-Saxon density
\begin{equation}
n\left(r\right)=\frac{n_{0}}{1+{\rm exp}\left(r-R\left(1+\beta_{2}Y_{20}+\beta_{4}Y_{40}\right)\right)/a}.
\end{equation}
The parameters used for \nucCu, \nucAu, and \nucU~ nuclei are listed in Table~\ref{tab:nuclpar}. 
In the collision, the principal axis of the deformed nucleus is randomly rotated in polar and azimuthal planes. The orientation of the deformed nucleus relative to the beam axis has a direct influence on centrality and eccentricity of the collisions at a given fixed value of impact parameter $b$. The effects of the deformation are most important~\cite{Filip:2007tj,Filip:2009zz,Filip:2010zz} for the central collisions of strongly deformed nuclei, such as \nucU. 

The short range NN repulsion is simulated in the {\tt GLISSANDO} nuclear distributions via the introduction of nucleon-nucleon expulsion distance $d$. The centers of nucleons in each nucleus cannot be closer to each other than $d$, which simulates the hard-core repulsion in the nuclear potential. The used value $d=0.9$~fm reproduces accurately~\cite{Broniowski:2010jd} the effects of the central NN correlations implemented in a more exact manner in  Refs.~\cite{Alvioli:2009ab,Alvioli:2010yk}.  

In the wounded-nucleon model the key entity is the NN collision profile, $p(b)$, defined as the probability of inelastic nucleon-nucleon collision at the impact parameter $b$.  Most of the Glauber Monte Carlo generators on the market use, for simplicity, the hard-sphere wounding profile, i.e., the collision occurs if $b<\sqrt{\sigma_{\rm w}/(2\pi)}$ for the wounded, and $b<\sqrt{\sigma_{\rm bin}/(2\pi)}$ for the binary collisions.
As shown in~\cite{Bialas:2006qf}, the NN wounding profile in the form of a combination of Gaussians can precisely reproduce the CERN ISR experimental data~\cite{Bohm:1974tv,Nagy:1978iw,Amaldi:1979kd,Amos:1985wx,Breakstone:1984te}  
on the total and elastic differential $p+p$ cross section. Here we use a single Gaussian form, which for the studied heavy-ion observables is accurate enough~\cite{Rybczynski:2011wv},
\begin{equation}
  p(b)=Ae^{-\pi A b^{2}/ \sigma_{\rm w}}.
\end{equation}
The parameter $A$ depends weekly on the collision energy and we use $A=0.92$ for our studies~\cite{Rybczynski:2011wv}.  
The realistic Gaussian wounding profile affects important observables in a noticeable way, as shown in~\cite{Rybczynski:2011wv}. Namely, it reduces the eccentricity parameters as well as multiplicity fluctuations. The effects are at the level of 10-20\% compared to the hard-sphere profile. Physically, the effects enter from the fact that the Gaussian profile has a tail extending to large values of $b$, thus nucleons staying far away from the collision center may collide with a non-zero probability. 

For various comparisons presented in this paper, we are using three different models as implemented in {\tt GLISSANDO}~\cite{Broniowski:2007nz}, namely the wounded-nucleon model, the mixed model~\cite{Kharzeev:2000ph}, and the hot-spot model~\cite{Gyulassy:1996br,Broniowski:2007nz} with overlaid $\Gamma$ distribution.
Within the Glauber approach, during the first stage of the collision individual interactions between the nucleons deposit transverse entropy (or energy). These elementary processes, stemming from wounded nucleons or binary collisions, are termed {\em sources}. A weight called relative deposited strength (RDS) is assigned to each source. The distribution takes the form 
\begin{eqnarray}
f(RDS) = (1-\alpha)N_{\rb{W}}/2 f_{\rb W}+\alpha N_{\rb{bin}} f_{\rb{bin}},
\end{eqnarray}
where $\alpha$ is the parameter controlling the relative weight of the wounded to binary sources, $N_{\rb{W}}$ and $N_{\rb{bin}}$ are the numbers of the wounded nucleons and binary collisions, while $f_{\rb{W}}$ and $f_{\rb{bin}}$ are the statistical distributions of the RDS generated by the wounded nucleons and binary collisions, respectively. We assume the normalization $\int du f_{\rb{W}}(u) = \int du f_{\rb{bin}}(u)=1$.

For the wounded-nucleon model with no superimposed distribution $\alpha=1$ and $f_{\rb W}(u)=\delta(u-1)$, i.e. the strengths of each source is equal. This provides a lower limit for the amount of initial fluctuations in the entropy-production mechanism. 

For the mixed model investigated here we assume $\alpha=0.145$ (the value fitting the multiplicity distributions 
at the highest RHIC collision energy~\cite{Back:2001xy,Back:2004dy}), and $f_{\rb W}(u)=f_{\rb bin}(u)=\delta(u-1)$. This is a popular choice in many other simulations.

As mentioned in the Introduction, realistically, the distribution of sources in the transverse plane should be convoluted with a statistical distribution. This convolution simulates the dispersion in the generated transverse entropy. 
Since the meaning of hot-spots is somewhat different in various theoretical models, let us explain in detail our implementation~\cite{Broniowski:2007nz}. In our {\em hot-spot model} a binary collision is generated according to the standard criterion, but is accepted with the probability \mbox{$\sigma_{\rm bin}/\sigma_{\rm w}$}, whereas the RDS of \mbox{$\alpha\sigma_{\rm w}/\sigma_{\rm bin}$} is assigned to the hot-spot position in the transverse plane.  
We use $\sigma_{\rm bin}=2$~mb, which means that the hot-spots occur rarely (in 5\% of binary collisions), but with a large weight (about 20 times larger than the wounded sources). The average weight of events is from construction equal to $(1-\alpha)N_{\rb{W}}/2+\alpha N_{\rb{bin}}$, the same as in the mixed model, but it can fluctuate considerably from event to event depending on how many hot-spots are created. In addition, we overlay the $\Gamma$ distribution on top of the wounded nucleons and hot-spots, with the distributions $f_{\rb W}(u)=f_{\rb bin}(u)=\Gamma(u, \kappa)$, where
\begin{eqnarray}
\Gamma(u,\kappa)=\frac{u^{\kappa-1}\kappa^\kappa \exp(-\kappa u)}{\Gamma(\kappa)}, \;\; u \in [0,\infty). \label{gamma}
\end{eqnarray} 
The $\Gamma$ distribution gives $\langle u \rangle=1$ and ${\rm var}(u)=1/{\kappa}$. In our simulations we use $\kappa=2$.

We have analyzed $10^{7}$ minimum bias \UUcol~collisions and the same number of \CuAucol~events generated by {\tt GLISSANDO}. Since the minimum bias simulations contain very limited statistics of the very central collisions, we have prepared additional samples for this case. All events were generated with the total inelastic NN cross section $\sigma_{\rm w}$=42 mb, which is the value assumed for the energy \roots= 200 GeV. 

The RDS determines the centrality classes and its meaning depends on the model. The resulting centrality classes are listed in Tables~\ref{tab:cent_CuAu} and \ref{tab:cent_UU}. From construction, the centrality classes are essentially the same for the mixed and the hot-spot+$\Gamma$ models, as the overlaid distributions $f_{\rb W}(u)=f_{\rb bin}(u)$ have the mean value set to one.

\begin{figure} 
\begin{center} 
\includegraphics[width=.55\textwidth]{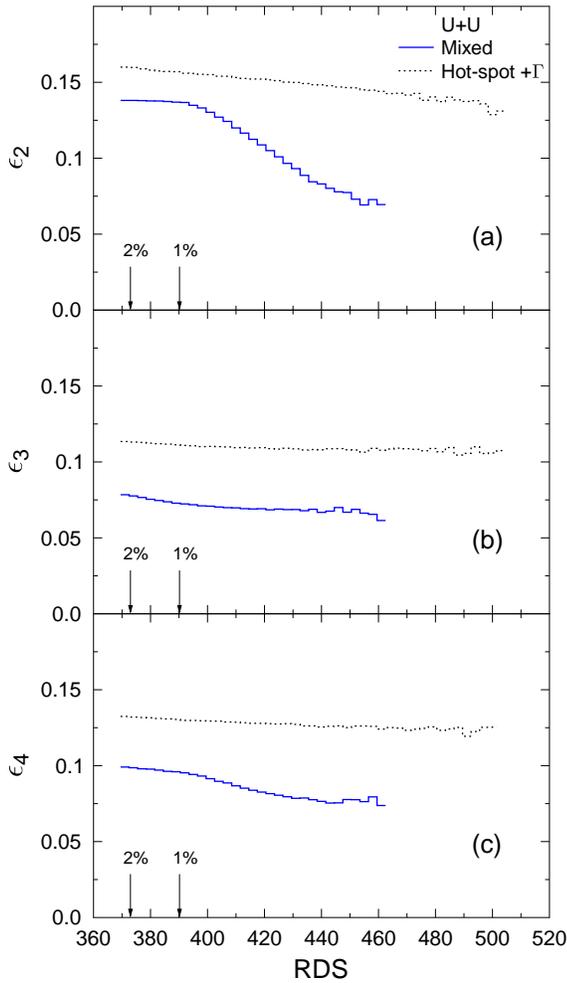}
\end{center}
\caption{(Color online) The event-average harmonic eccentricities $\epsilon_n$ ($n=2,3,4$) for the most central U+U collisions, plotted as function of the relative deposited strength (RDS) for the mixed model and the hot-spot+$\Gamma$ model,  $\sigma_{\rm w}=42$~mb with the Gaussian wounding profile.
\label{fig_epsn_UU_cent_comp_RDS}} 
\end{figure}

\section{Shape deformation}
\label{sect:hi-ord-fourier}

In this Section we investigate the Fourier components of the azimuthal distribution of sources in the transverse plane, defined as  
\begin{equation}
 \epsilon_{n} = \frac{\sqrt{(\sum_{i} r^{n}_{i}{ \rm cos}[n(\phi_{i}\!-\!\Phi_{n})])^{2}\!+\!(\sum_{i}r^{n}_{i}{ \rm sin}[n(\phi_{i}\!-\!\Phi_{n})])^{2}} } {\sum_{i} r^{n}_{i}}
\end{equation}
where $i$ runs over the number of sources in each event, $r_{i}$ is the distance of the source from the center of mass of the fireball, and $\phi_{i}$ is its azimuthal angle. The weight associated to each source is proportional to $r^n$~\cite{Broniowski:2007ft}, which makes the higher moments more sensitive to sources further away from the center.

The angle $\Phi_{n}$ is adjusted in each event in such a way as to maximize $\epsilon_{n}$, which gives a condition 
\begin{equation}
{\rm tan}~n \Phi_{n} = \frac{\sum_{i}r^{n}_{i}{~\rm sin}(n \phi_{i})}{\sum_{i}r^{n}_{i}{~\rm cos}(n \phi_{i})}. \label{eq:axes}
\end{equation}
The angle $\Phi_{n}$ is the phase of the principal axis of $n$th order azimuthal Fourier component.

\begin{figure} 
\begin{center} 
\includegraphics[width=.4\textwidth]{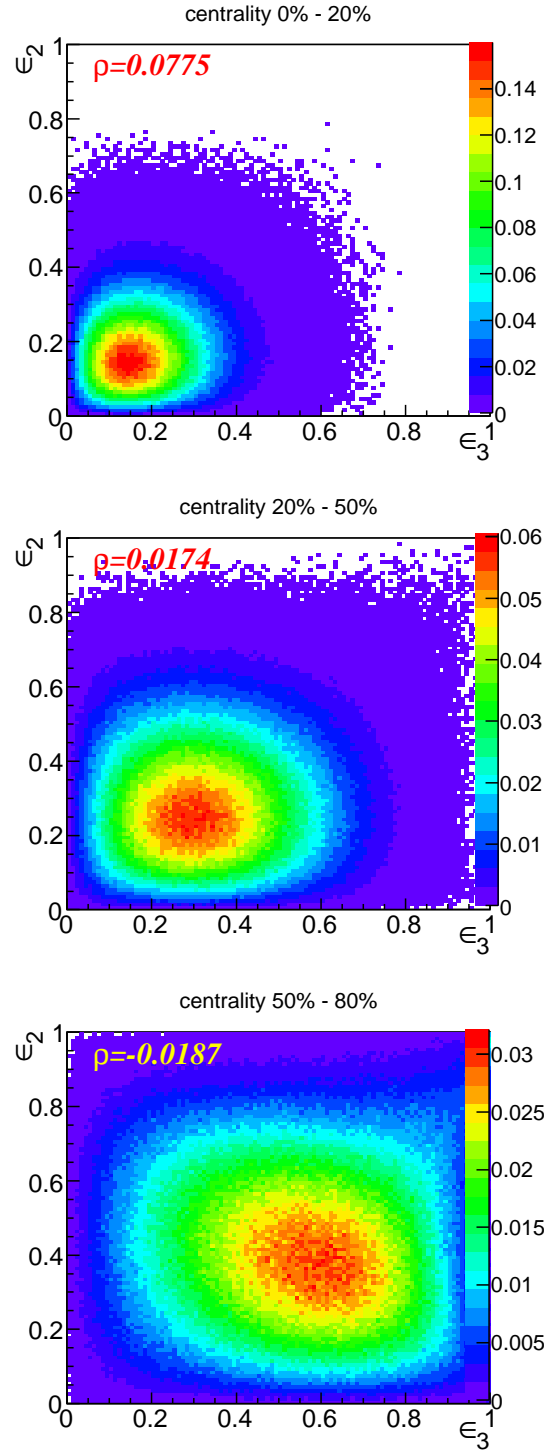}
\end{center}
\caption{ (Color online) Two-dimensional event-by-event distribution plot of $\epsilon_2$ and $\epsilon_3$ for the Cu+Au collisions at three centrality classes, the hot-spot$+\Gamma$ model, $\sigma_{\rm w}=42$~mb with the Gaussian wounding profile. The value of the correlation coefficient is given in the upper left corners.
\label{ep2_vs_ep3_42_CuAu_hotspot_rn_gwp}} 
\end{figure} 

\begin{figure} 
\begin{center} 
\includegraphics[width=.4\textwidth]{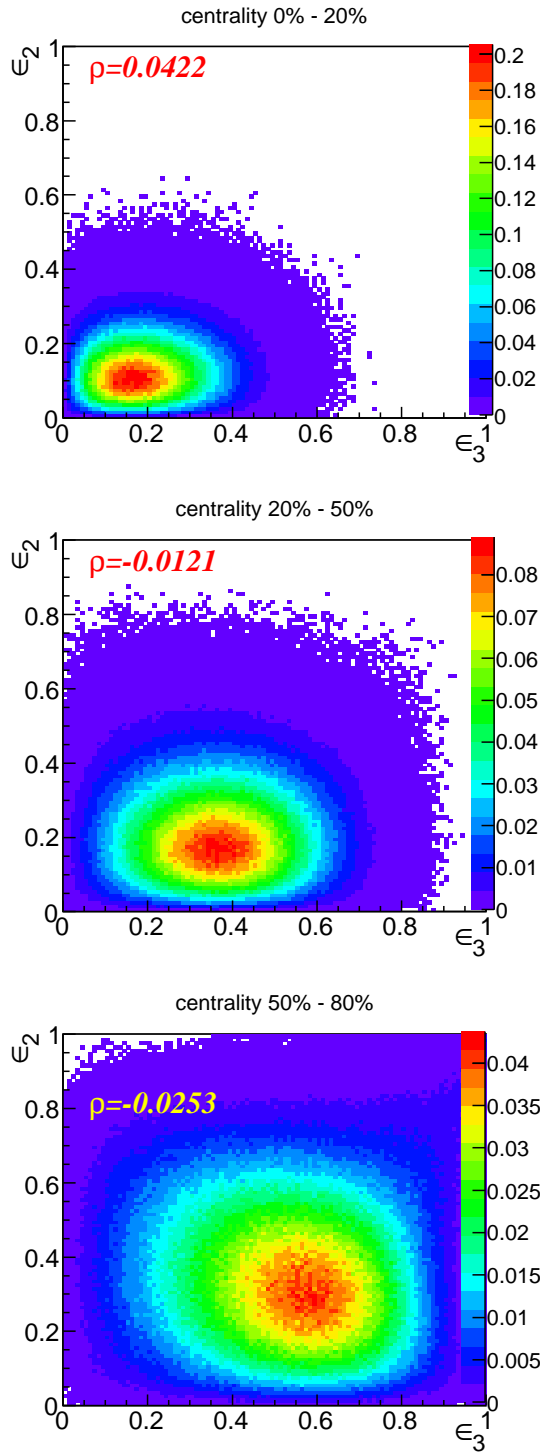}
\end{center}
\caption{(Color online) Same as Fig.~\ref{ep2_vs_ep3_42_CuAu_hotspot_rn_gwp} for the U+U collisions.
\label{ep2_vs_ep3_42_UU_hotspot_rn_gwp}} 
\end{figure}

In Fig.~\ref{fig_epsn_mixed_gwp_rn_comp} we test the effects of nuclear deformation. We display the dependence of event-average $\epsilon_{n}$ ($n$=2,3,4) on the number of wounded nucleons, $N_{W}$, in the mixed model. The figures show two sets of curves, solid and dashed, corresponding to the case with and without deformation, respectively. The largest difference is observed for the eccentricity $\epsilon_{2}$ for the U+U collisions at highest centralities. For other cases the influence of deformation is small (we have checked this up to $\epsilon_{6}$). 

For \CuAucol~we can see several rather weak effects connected with the deformation of the colliding nuclei. First, $\epsilon_2$ is increased in most central events, decreased in semi-central collisions, and in addition the maximum value of $N_{W}$ is larger. These effects can be inferred geometrically from the fact that we collide small prolate \nucCu~nucleus and a much bigger \nucAu~nucleus with the oblate deformation. The triangular deformation parameter $\epsilon_{3}$ increases due to deformation for the mid-central events.

The effects observed in the very central \UUcol~collisions of deformed nuclei are further investigated in Fig.~\ref{fig_epsn_UU_cent_comp_RDS}, where we show $\epsilon_{n}$ ($n$=2,3,4) as a function of RDS. We compare the mixed model and the hot-spot + $\Gamma$ model in order to visualize the influence of the initial fluctuations. We note large qualitative differences between the two considered models. In the mixed model we see a ``knee'' structure, as advocated in~\cite{Voloshin:2010ut}, in $\epsilon_2$ around $RDS=400$. Such a behavior is a consequence 
of the tip-tip orientation of the prolate \nucU~nuclei. In the most central events the initial eccentricity is reduced because the transverse profile of longitudinally oriented \nucU~nuclei is spherical. However, in the hot-spot + $\Gamma$ the knee structure disappears due to fluctuations from the hot-spots and the overlaid distribution. 
One thus observes that the hot-spot model, which assumes the production of large amount of entropy in rare semi-hard binary collisions (hot-spots), hides the purely geometrical effects in $\epsilon_{2}$. The events from the hot-spot model exhibit much higher values of all $\epsilon_{n}$ ($n$=2,3,4). The increase is around 30-50\% or even higher for the very central collisions. As shown in~\cite{Rybczynski:2011wv,Filip:2009zz}, even for collisions of spherical nuclei the event-by-event fluctuations of eccentricity $\epsilon_{2}$ are quite large. The deformation of colliding nuclei can slightly increase these fluctuations, as the similar number of sources can appear in collisions of nuclei with different orientations in the azimuthal plane.

\begin{figure} 
\begin{center}
\includegraphics[width=.5\textwidth]{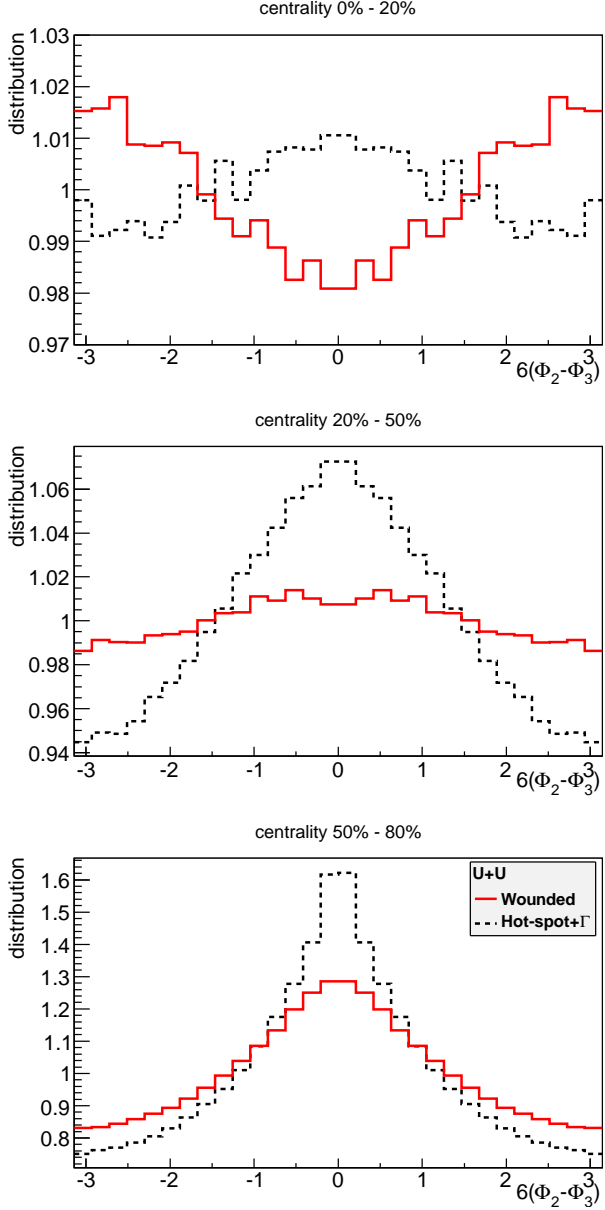}
\end{center}
\caption{The wounded-nucleon (solid) and hot-spot$+\Gamma$ (dashed) model prediction for the distribution of $6(\Phi_{2}-\Phi_{3})$ in Cu+Au collisions.
\label{phi_cent_42_CuAu_rn_gwp}} 
\end{figure} 

\begin{figure}
\begin{center} 
\includegraphics[width=.5\textwidth]{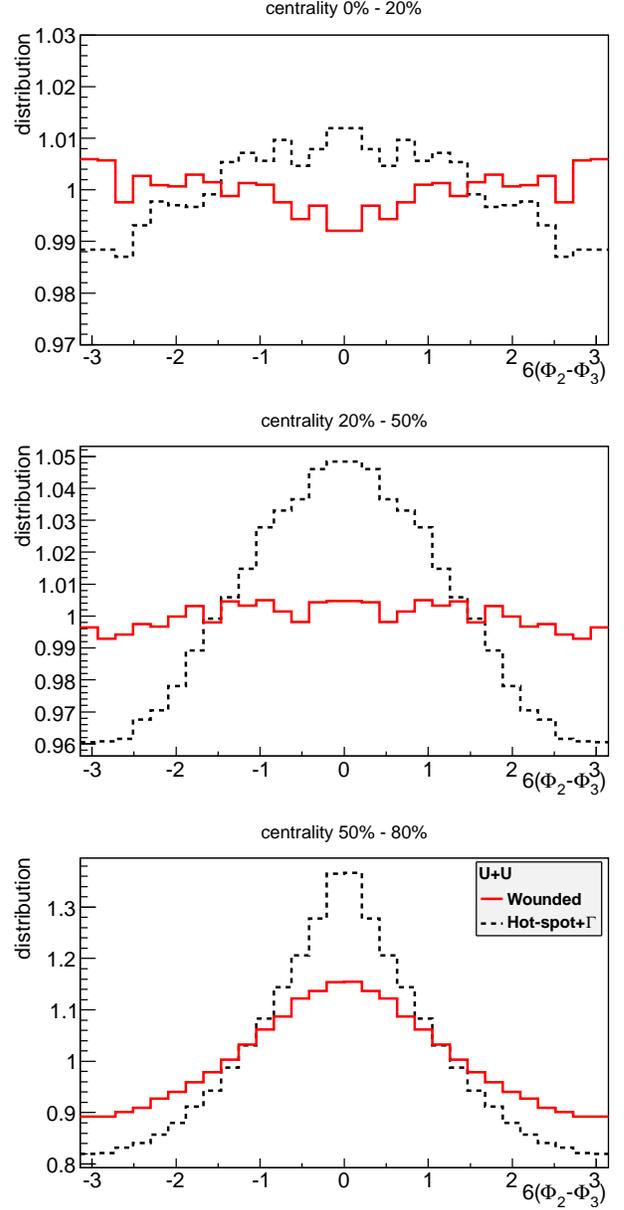}
\end{center}
\caption{Same as Fig.~\ref{phi_cent_42_CuAu_rn_gwp} for the U+U collisions.
\label{phi_cent_42_UU_rn_gwp}} 
\end{figure}

\begin{figure}
\begin{center}
\includegraphics[width=.25\textwidth]{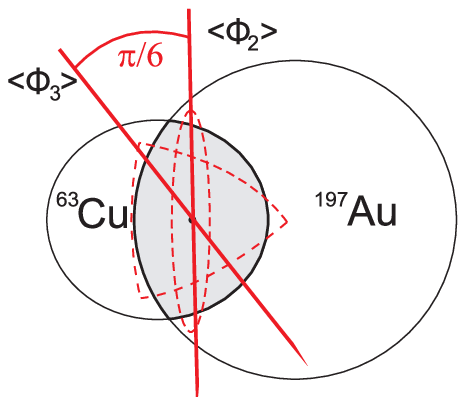}
\end{center}
\caption{(Color online) A cartoon of the geometry of a mid-central collision of asymmetric nuclei. The smaller nucleus 
has a larger curvature at the geometric boundary than the larger nucleus, thus the overlap region has a shape of an asymmetric almond, hence the axes of the elliptic and triangular deformations form an angle of $\sim \pi/6$.  
\label{fig:geom}} 
\end{figure}

\section{Event-plane correlations}
\label{sect:ev-pl-cor}

Next, we show the analysis of the correlation between different Fourier components $\epsilon_{n}$ of the azimuthal 
distribution of sources as well as the correlation between the direction $\Phi_{n}$ of the principal axes.  Figures~\ref{ep2_vs_ep3_42_CuAu_hotspot_rn_gwp} 
and \ref{ep2_vs_ep3_42_UU_hotspot_rn_gwp} present 2D correlation plots for $\epsilon_{2}$ and
$\epsilon_{3}$ in three selected centrality ranges of the \CuAucol~ and \UUcol~ collisions. They show that the correlation between $\epsilon_{2}$ and  $\epsilon_{3}$ is weak and does not change significantly with centrality. 
The correlation is defined in the standard way as
\begin{eqnarray}
\rho = \langle (\epsilon_2 - \langle \epsilon_2 \rangle) (\epsilon_2 - \langle \epsilon_2 \rangle) \rangle.
\end{eqnarray}
The largest correlation ($\rho\simeq 8\%$) is observed in the central (0-20\%) \CuAucol~ events. In peripheral collisions (50-80\%) $\epsilon_{2}$ and $\epsilon_{3}$ are slightly anti-correlated. From Figs.~\ref{ep2_vs_ep3_42_CuAu_hotspot_rn_gwp} and \ref{ep2_vs_ep3_42_UU_hotspot_rn_gwp} we note that the small correlation may be induced by the finite size of the space ($0 \le \epsilon_n \le 1$) rather than any dynamics. 

Next, we discuss the correlation between directions $\Phi_{n}$ of principal axes (\ref{eq:axes}). Most interesting is the correlation of the lowest rank even and odd axes, i.e., $\Phi_{2}$ and $\Phi_{3}$. 
A correlation of these axes is clearly seen in Figs.~\ref{phi_cent_42_CuAu_rn_gwp} and \ref{phi_cent_42_UU_rn_gwp}, 
where the distribution of the angle $6(\Phi_{2} - \Phi_{3})$ is shown in three large centrality bins for the two colliding systems and two models: the wounded-nucleon model and the hot-spot model. We note that at low centralities these models yield qualitatively different results.

\begin{figure*}
\begin{center}
\includegraphics[width=1.2\textwidth]{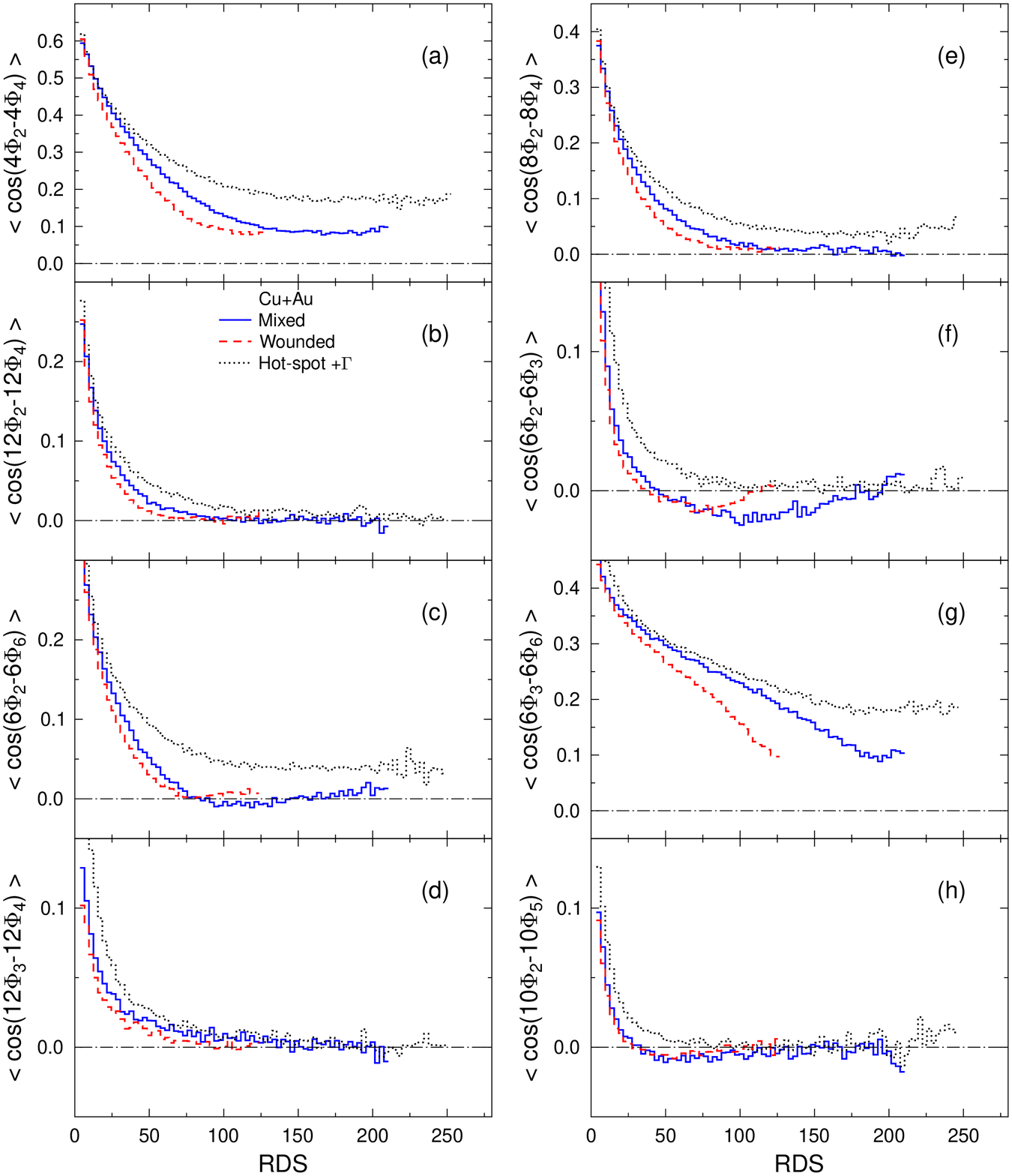}
\end{center}
\caption{(Color online) The event-plane correlations measures $\langle {\rm cos} [k(\Phi_n-\Phi_m)]\rangle $ for the Cu+Au collisions. We compare the predictions of the mixed model (solid line), the wounded-nucleon model (dashed line), and the hot-spot$+\Gamma$ model (dotted line).
\label{fig_phi_CuAu_mixed_gwp_rn_comp}} 
\end{figure*} 

\begin{figure*}
\begin{center}
\includegraphics[width=1.2\textwidth]{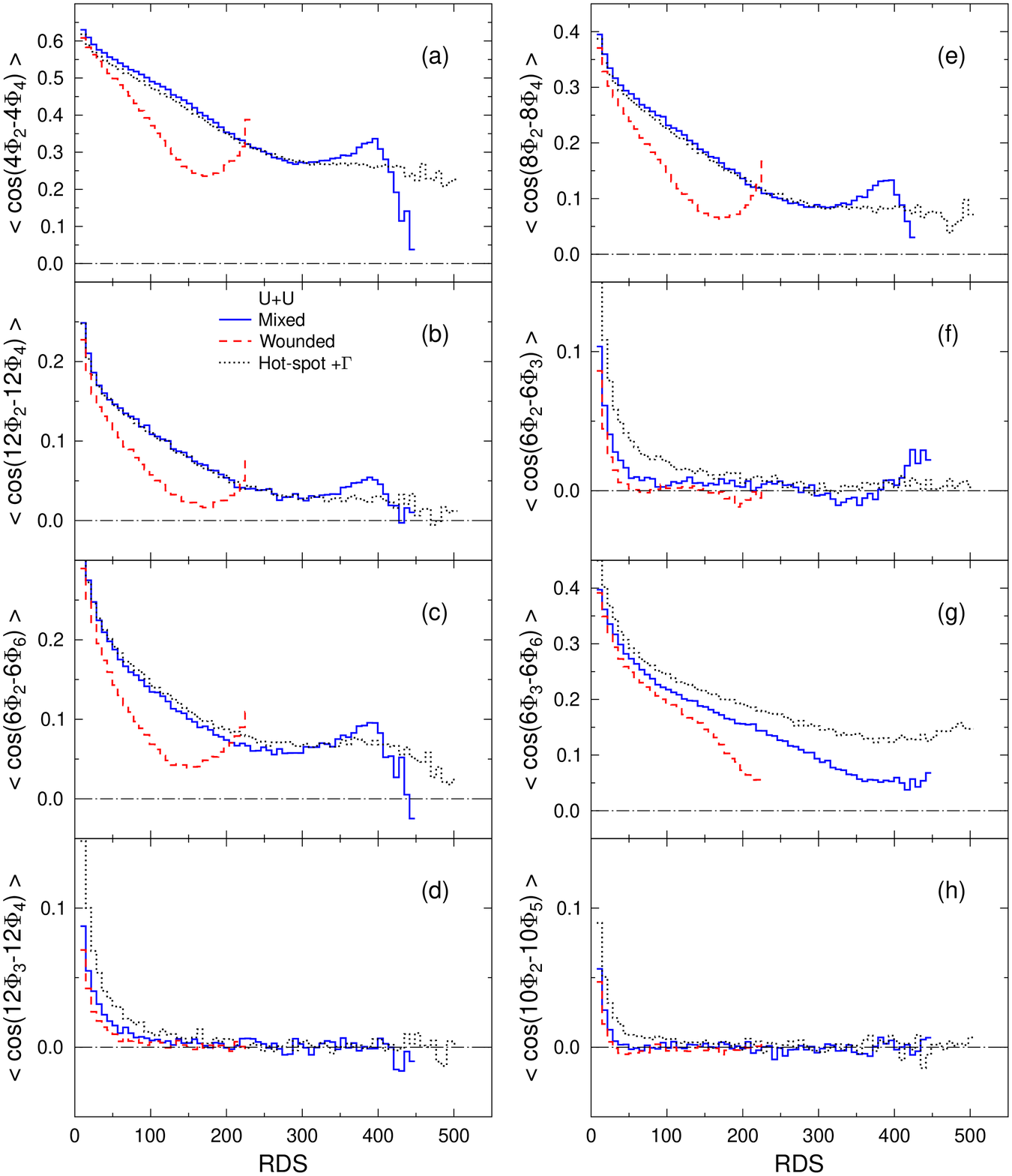}
\end{center}
\caption{(Color online) Same as Fig.~\ref{fig_phi_CuAu_mixed_gwp_rn_comp} for the U+U minimum collisions. 
\label{fig_phi_UU_mixed_gwp_rn_comp}} 
\end{figure*}

For the wounded-nucleon model these distributions exhibit a minimum at $\Phi_{2}$ - $\Phi_{3}=0$ in central collisions (centrality 0-20\%). This can be understood as follows. The angle $\Phi_{2}$ fluctuates around the direction perpendicular to the reaction plane. The triangularity angle $\Phi_{3}$ naively should be completely random. However, since in the colliding systems there are more nucleons distributed in the direction parallel to the reaction plane than perpendicular, there is a higher probability that one of the corners of the triangle points in the direction parallel to the reaction plane. Then the relative angle between the average values of $\Phi_{2}$ and $\Phi_{3}$ is $\pi/6$ (see Fig.~\ref{fig:geom}), and $6(\Phi_{2}$ - $\Phi_{3})$ shows maxima at $\pm \pi$. The effect is stronger for the asymmetric Cu+Au system, as in this case the difference in the radii of the nuclei forms on the average the arrangement of Fig.~\ref{fig:geom}). At larger centralities the nuclei collide more peripherally, 
with fewer participants. In that case $\Phi_{3}$ follows closer and closer $\Phi_{2}$ and the distribution of Figs.~\ref{phi_cent_42_CuAu_rn_gwp} and \ref{phi_cent_42_UU_rn_gwp} become more sharply peaked at zero. 

In the hot-spot+$\Gamma$ model (dashed lines in Figs.~\ref{phi_cent_42_CuAu_rn_gwp} and \ref{phi_cent_42_UU_rn_gwp})
the situation is different. The presence of a strong hot-spot source collimates $\Phi_{2}$ and $\Phi_{3}$ even 
at low centralities, hence the maximum of the distribution is at $\Phi_{2}$ - $\Phi_{3} = 0$.

The observed qualitative difference of the distribution of $(\Phi_{2} - \Phi_{3})$ might be used to discriminate between different models of the early phase, and in particular, the degree of the initial fluctuations. However, we should keep in mind that the angles $\Phi_n$ are not observable. The observable angles are the event-plane angles, $\Psi_n$, determined from measuring the momenta of produced particles. Nevertheless, it is expected that the correlations between the $n=2$ and $3$ axes are carried over by event-by-event hydrodynamics from the initial state to the final particle distributions~\cite{Qiu:2011iv}. The effects shown in Figs.~\ref{phi_cent_42_CuAu_rn_gwp} and \ref{phi_cent_42_UU_rn_gwp} are small for the central and mid-central collisions (the departures from unity are at the level of 1-2\%), but since the experimental resolution is very good, they should be within the 
experimental reach. 

A convenient measure of the above-discussed effect is 
\begin{eqnarray}
\langle \rm cos[k(\Phi_{n}-\Phi_{m})] \rangle, 
\end{eqnarray}
where $k$ is the least common multiple of $n$ and $m$ multiplied by small natural number, and averaging is over the events in a given class. Such two-plane correlators are sometimes calculated for experimental data, and their linear combinations provide the three-plane correlators which carry additional information not accessible through two-plane correlators. When averaged with the distributions such as in Figs.~\ref{phi_cent_42_CuAu_rn_gwp} and \ref{phi_cent_42_UU_rn_gwp}, the value becomes negative when the distributions are concave, and positive when they are convex. Thus the more collimated axes, the higher $\langle \rm cos[k(\Phi_{n}-\Phi_{m})] \rangle$ is.  

The centrality dependence of $\langle \rm cos[k(\Phi_{n}-\Phi_{m})] \rangle$ for different choices of $n$ and $m$ 
in \CuAucol~ and \UUcol~collisions is shown in Figs.~\ref{fig_phi_CuAu_mixed_gwp_rn_comp} and Fig.~\ref{fig_phi_UU_mixed_gwp_rn_comp}. The presented results suggest that the correlation between $\Phi_{n}$ and $\Phi_{m}$ is strong in peripheral collisions for all measured $n$ and $m$. This simply follows from the fact that formally in the limit of very peripheral collisions we have just two sources (when there is only one source we cannot 
determine the axes and the event is excluded from the sample). In that case all axes $\Phi_n$ are parallel to one another (and also $\epsilon_n=1$). One can see that the correlation between $\Phi_{2}$ and $\Phi_{4}$, as well as $\Phi_{3}$ and $\Phi_{6}$ is quite strong also for the mid-central events. The shape of the distributions for \CuAucol~ and \UUcol~ collisions significantly differs only for the correlators of $\Phi_{2}$ and $\Phi_{4}$. We note a knee structure in the centrality dependence of the $\Phi_{2}$ and $\Phi_{4}$ correlation in the most central \UUcol~collisions for the mixed model. Similarly to the case of $\epsilon_2$ of Sec.~\ref{sect:hi-ord-fourier}, the structure disappears when the hot-spot+$\Gamma$ model is used, showing sensitivity to the presence of the initial fluctuations. 

The correlation of $\Phi_{2}$ and $\Phi_{4}$ in \CuAucol~collisions increases for central and mid-central events 
for the hot-spot model in comparison to the mixed model. The increase is most significant for $\rm cos(4(\Phi_{2}-\Phi_{4}))$. For \UUcol~collisions the hot-spot model does not change the strength of the $\Phi_{2} - \Phi_{4}$ correlation but the knee structure observed in the most central events disappears.
The correlation of $\Phi_{3}$ and $\Phi_{6}$ is increased in the central \CuAucol~and \UUcol~collisions for the hot-spot model.  

The quantities in Figs.~\ref{fig_phi_CuAu_mixed_gwp_rn_comp} and~\ref{fig_phi_UU_mixed_gwp_rn_comp} are plotted as functions of the RDS. The corresponding values of centralities can be obtained from Tables~\ref{tab:cent_CuAu} and \ref{tab:cent_UU}.

\section{Conclusions}
\label{sect:concl}

The main point of this paper is that the initial fluctuations may qualitatively change the behavior of various 
measures of the initial geometry (eccentricity parameters, event-plane correlations). The studies of Cu+Au and U+U
systems, recently analyzed at RHIC, clearly exhibit sensitivity to the choice of the selected variant of the Glauber-like model of the initial phase. In our study we have used {\tt GLISSANDO} simulations with deformed nuclei and the realistic Gaussian wounding profile for the NN collisions. We have shown that the model with a higher degree of fluctuations, such as the hot-spot+$\Gamma$ model, may easily hide the subtle features of the initial geometry, such as those reflecting the deformation of the colliding nuclei. Conversely, a sufficiently precise measurements of the event-plane correlators may indicate a particular model of the early phase, provided the correlations are not largely altered by the intermediate evolution of the system from the early phase to the freeze-out.

In view of our results, one may reconcile the prediction of~\cite{Voloshin:2010ut} for the knee shape in the dependence of the eccentricity parameter on the number of produced particles in the most central U+U collisions. No knee in the experimental data would hint to an initial model with more fluctuations.

\begin{acknowledgments}

This work was supported by the Polish Ministry of Science and Higher Education under Grants No. N N202 263438, 
N N202 288638, and National Science Centre, grant DEC-2011/01/D/ST2/00772.                                      

\end{acknowledgments}

\end{document}